\documentclass{optica-article}
\journal{opticajournal} % for journals or Optica Open
\articletype{Research Article}

\usepackage[utf8]{inputenc}
\usepackage{mathtools}
\usepackage{xcolor}
\usepackage{graphicx}
\usepackage{verbatim}

\usepackage{newfloat,algcompatible} 
\usepackage{algorithm}
\usepackage{algpseudocode}
\usepackage{braket}
\usepackage{bm}
\usepackage{amsmath}
\usepackage{nicematrix}
\usepackage{float}

\newcommand{\comm}[1]{\textcolor{red}}

\begin{document}

\title{Translation-based structured illumination microscopy via generalized Richardson-Lucy method} % Force line breaks with \\

\author{Valentina Capalbo,\authormark{1,2,*} 
Damiana Battaglini,\authormark{3,4} 
Marialaura Petroni,\authormark{3,4} 
Giancarlo Ruocco,\authormark{1,5} 
Marco Leonetti,\authormark{6,1,2}}

\address{\authormark{1} Center for Life Nano- \& Neuro-Science, Italian Institute of Technology, Rome, Italy}

\address{\authormark{2} D-Tails s.r.l. BCorp, Via Agrigento 4a 4b, 00161 Rome, Italy}

\address{\authormark{3} Department of Molecular Medicine, Sapienza University of Rome, I-00185 Rome, Italy}

\address{\authormark{4} Istituto Pasteur-Fondazione Cenci Bolognetti, Rome, Italy.}

\address{\authormark{5} Department of Physics, Sapienza University of Rome, I-00185 Rome, Italy}

\address{\authormark{6} Institute of Nanotechnology of the National Research Council of Italy, CNR-NANOTEC, Rome Unit, Piazzale A. Moro 5, I-00185 Rome, Italy}

\email{\authormark{*} valentina.capalbo@iit.it}%% email address is required; see note below about the corresponding author designation

% use {asbstract*} to suppress the copyright line. Copyright information will be added in production

\begin{abstract*}
Structured illumination microscopy (SIM) can achieve a $2\times$ resolution enhancement beyond the classical diffraction limit by employing illumination translations with respect to the object. This method has also been successfully implemented in a ``blind'' configuration, i.e., with unknown illumination patterns, allowing for more relaxed constraints on the control of illumination delivery. Here, we present a similar approach using a novel super-resolution algorithm that employs a generalized version of the popular Richardson-Lucy algorithm, alongside an optimized and customized optical setup. Both numerical and experimental validations demonstrate that our technique exhibits high noise resilience. Moreover, by implementing random translations instead of ``ordered'' ones, noise-related artifacts are reduced. These advancements enable wide-field super-resolved imaging with significantly reduced optical complexity.
\end{abstract*}

%%%%%%%%%%%%%%%%%%%%%%%%%%  body  %%%%%%%%%%%%%%%%%%%%%%%%%%
\section{Introduction}
\label{sec:introduction}
The optical objective is the keystone of fluorescence microscopy, defining both the field of view $\mathcal{F}$ (i.e., the extent of the imaged area diameter) and the optical resolution $\mathcal{R}$ (i.e., the minimal length scale which can be properly distinguished and analyzed). The total amount of information extracted from an image is thus $\mathcal{I}=\frac{\mathcal{F}^2}{\mathcal{R}^2}$. A prevailing trend in biology is to integrate information across multiple scales \cite{he2023high, shi2019multi, oh2014mesoscale} in a cohesive manner to uncover mechanisms underlying organ-scale biological functions, thus novel techniques merging ultra-wide fields with enhanced resolution are highly sought. While mechanical stage scanning and image stitching can be used to extend the effective field of view for a given objective, these approaches can be time-consuming and prone to alignment errors or motion artifacts. Therefore, methods that enable simultaneous improvements in resolution and native wide-field acquisition are particularly desirable. Notably, it is still not possible to enlarge $\mathcal{F}$ directly for any given physical objective; however, it is possible to act on $\mathcal{R}$ resorting to super-resolution (SR) techniques that enable imaging beyond the classical diffraction limit \cite{abbe1878,rayleigh1896}. Several strategies have been developed to achieve the resolution enhancement \cite{schermelleh2010}, by exploiting the different response of the fluorophores to excitation or molecular localization methods. Examples include depletion-based techniques, such as stimulated emission depletion (STED, \cite{hell1994breaking,klar1999sted}), and stochastic activation methods, such as photo-activated localization microscopy (PALM, \cite{hess2006palm}) or stochastic optical reconstruction microscopy (STORM, \cite{rust2006storm}). However, these approaches require either high-power pulsed lasers or chemical engineering of fluorophores, making the effective resolution increase highly dependent on the experimental configuration. Notably, alternative nonlinear optical techniques have been recently developed for SR imaging, for example combining image scanning microscopy (ISM) with STED \cite{tortarolo2022}, or multi-photon excitation of upconversion nanoparticles \cite{wang2024}.
One of the SR techniques that involves the smallest amount of experimental modifications compared to the wide-field fluorescence microscopy is the structured illumination microscopy (SIM, \cite{gustafsson2000surpassing}). SIM exploits structured light, i.e., illumination that is intentionally non-uniform or ``shaped" in a specific way, and its interaction with the sample through optical interference patterns. In this sense, standard SIM can be considered a linear SR technique, as it does not involve nonlinear optical effects and preserves the linear relationship between illumination and fluorescence emission. In fact, SIM improves resolution by shifting high-frequency information of the sample, that are in principle inaccesible, into the observable domain via interference fringes. In practice, the method involves exposing the sample to a sequence of known patterned illuminations with different phases and orientation, then applying specific algorithms to combine the acquired images into a super-resolved one.
Standard SIM thus relies on three key elements:
\begin{itemize}
    \item  Structured illumination pattern
    \item  Limited number (9-36) of image acquisitions, each corresponding to a different phase shift and orientation of the illumination pattern 
    \item  Complete knowledge of the illumination pattern  
\end{itemize}
SIM has been proposed in several declinations (see e.g. \cite{strohl2016frontiers,samanta2021,ma2021} for reviews), also including nonlinear approaches \cite{gustafsson2005}. Among the others, a more flexible version of SIM is its ``blind" variant (blind-SIM, \cite{mudry2012structured}) which does not require severe control and prior knowledge of the illumination pattern. Instead, it assumes that the illumination is, on average, uniform over the sample plane. In this approach, images are acquired using different realization of randomized speckle patterns \cite{labouesse2017joint,yeh2017structured,idier2018,leonetti2019scattering}. The key advantage of blind-SIM is its ability to generate illumination effortlessly, without requiring precise alignment between the user-designed illumination pattern and the pattern that actually reaches the sample. This ensures that aberrations in the illumination beamline do not degrade the super-resolved image reconstruction. However, the primary drawback is the significantly higher number of required acquisitions (10 to 100 times more than standard SIM), along with increased computational costs to restore both the super-resolved image of the sample and the unknown illumination. Essentially, blind-SIM trades the prior knowledge of illumination patterns for a larger number of acquisitions.
To summarize, the three main components of blind-SIM are:
\begin{itemize}
    \item  Structured illumination pattern, randomized at each acquisition
    \item  Large number ($\sim 50-600$) of image acquisitions, each relative to a different illumination pattern
    \item  Statistically homogeneous illumination of the sample
\end{itemize}
Among the various reconstruction algorithms developed for SIM \cite{chen2023superresolution}, the advent of deep learning has significantly improved results by enhancing image quality, reducing noise, and accelerating processing speed (see e.g. \cite{wang2019,ling2020,xypakis2023,wu2024}).
Notably, a recently developed technique \cite{xypakis2022} has, for the first time, applied neural networks to blind-SIM, demonstrating that it is possible to both reduce computational time and enhance resolution. 
In this context, a new paradigm emerged with the seminal work presented in \cite{yeh2019computational,yeh2019speckle}, which introduced a novel approach known as computational SIM (C-SIM). C-SIM employs a constant unknown illumination that is slightly translated across the field of view at each acquisition. This method only needs an extremely steady illumination pattern throughout the acquisition/translation process and the knowledge of the translation amplitudes. By using advanced computational algorithms, C-SIM enables the reconstruction of high-resolution images of the sample fluorescence and the unknown illumination while simultaneously optimizing system imperfections, such as aberrations and deviations in the translation trajectory.
The main requirements of C-SIM can be summarized as follows:
\begin{itemize}
    \item  Structured illumination pattern, constant along all the acquisitions
    \item  A sequence of image acquisitions ($\sim 50-600$) corresponding to each translation of the illumination pattern
    \item  Exact knowledge of the translation amplitudes
\end{itemize}
In its original formulation, C-SIM assumed that the fluorescent sample remained fixed relative to the unknown translated illumination. However, this is not a strict requirement, as only the relative displacement between the illumination and the sample is required. For instance, a variation of this approach has recently been developed and tested in an ophthalmoscopic setting \cite{fusco2024}, where the stochastic movements of the sample---in this case, the intrinsic movements of the human eye---are exploited to develop a scan-less version of SIM, known as stochastically structured illumination microscopy (S$^2$IM). 
In this study, we present a novel optical technique that combines insights from S$^2$IM on the stochastic nature of the translations with a newly designed microscopy-specific experimental setup.
Our approach also introduces a new algorithm based on a customized generalization of the Richardson-Lucy (RL) method to retrieve both the high-resolution image of the sample and the unknown illumination pattern.
The RL algorithm \cite{richardson1972bayesian,lucy1974iterative} is one of the most widely used methods for reconstructing blurred and noisy images. Its key advantage lies in its straightforward iterative procedure for deconvolving measured data. Thanks to its adaptability to various imaging conditions, the RL algorithm is a valuable tool in different fields requiring high-quality image restoration, together with large field of view, such as microscopy, medical imaging, astronomy.
Originally developed for single-image deblurring with a known and constant point spread function (PSF) and Poisson noise corruption, the method has since been adapted for various experimental contexts. For instance, it has been used to process multiple images captured under varying illumination patterns \cite{ingaramo2014,strohl2015}, which has led to its application in SIM, also incorporating different noise contributions in the imaging process \cite{chakrova2016}.
Additionally, the method has been extended to support blind deconvolution, allowing for cases where the PSF is unknown or varies throughout the iterations \cite{fish1995,almeida2019}.
Here, we apply the RL algorithm by exploiting the statistical properties of stacked images, introducing a novel SR approach within a blind-SIM framework. In particular, we analyse results obtained using sample translations arranged in an ``ordered'' (square) lattice and in a ``disordered'' configuration, respectively, with an unknown illumination pattern. Then, we compare the performance of our new algorithm with the one used in C-SIM.

\section{Methods}

\subsection{Translated images and unknown structured illumination}
\label{sec:method}
\begin{figure}[H]
    \centering
    \includegraphics[width=10cm]{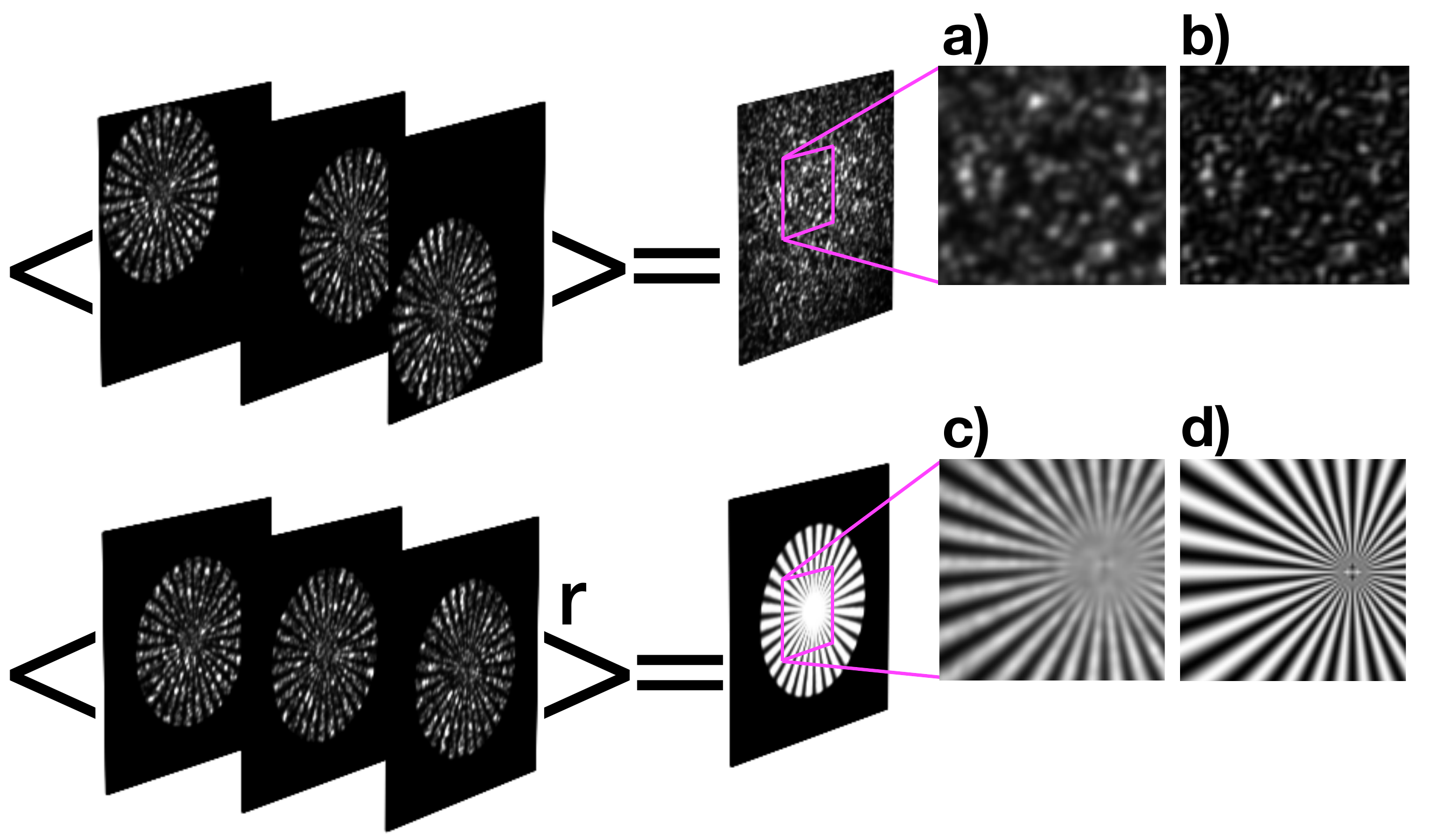}
    \caption{Illustration of \textit{simple averaging} and \textit{registered averaging} computed for a stack of 200 images.
    The images have a size of $571\times571$ pixels, with a pixel size of $1.38 \mu m$. The simulated object is a Siemens star of density $\rho=1+\cos30\theta$, while the illumination is a simulated speckle pattern with a numerical aperture of 0.035 and a wavelength of $0.605\,\mu m$. PSF blurring is simulated by convolving the images with a Gaussian kernel with FWHM$=8.64\,\mu m$.
    \textit{Top panel}: the average is computed over images acquired by translating the object while the illumination remains fixed. \textit{Bottom panel}: the average is computed on the registered images, where the object appears fixed and the illumination appears to shift. 
    The resulting blurred proxies of the pattern, $\bm{p}^0$ (see Eq.~(\ref{eq:simple averaging})), and the object, $\bm{\rho}^0$ (see Eq.~(\ref{eq:registered averaging})), are shown in the right insets a) and c), respectively. These are compared with their corresponding ground-truth images, displayed in insets b) and d).}
    \label{fig:averaging}
\end{figure}

Our idea for a new algorithm for SR with unknown illumination builds on two key properties of translated image stacks. These properties naturally arise from the way fluorescence microscopy signals are generated. Specifically, the image data, $\bm{D}(x,y)$, representing the intensity in the sample plane $(x,y)$, is the result of an element-wise multiplication between the fluorophore distribution, $\bm{\rho}(x,y)$, and the illumination pattern, $\bm{p}(x,y)$, followed by a blurring effect due to the finite optical resolution of the objective. This blurring is mathematically expressed as a convolution with the PSF of the objective, $\bm{h}(x,y)$. Then, the acquired image is given by
\begin{equation}
    \mathbf{D} \sim \left\{\bm{\rho}\cdot \bm{p} \right\}\otimes\bm{h}
\label{eq:data equation}
\end{equation}
where, to soften the notation, we omit the $(x,y)$ dependencies in all terms, while we indicate with $\cdot$ and $\otimes$, respectively, the element-wise multiplication and the convolution operator. Note that the approximation arises from disregarding the noise contribution. In our translation-based configuration, each acquisition $a=1, ..., N$ differs due to a translation of the object by a known amplitude. This is represented as
\begin{equation}
    \bm\rho_a=T( \bm\rho,\bm{\Delta}_a)
\label{eq:translation operator}
\end{equation}
where $T$ is the translation operator and $\bm{\Delta}_a=[\Delta x_a,\Delta y_a]$ defines the translation shift. Consequently, the expression for a set of images becomes
\begin{equation}
    \mathbf{D}_a \sim \left\{\bm{\rho}_a\cdot \bm{p} \right\}\otimes\bm{h}
\label{eq:data equation C-Sim}
\end{equation}
Note that, translating the sample relative to a fixed illumination is mathematically equivalent, in terms of the resulting image information, to translating the illumination across a fixed sample.
The full dataset of a set of acquisitions, i.e., the stack $\mathbf{\{D\}^N}$, can be simply averaged (\textit{``simple averaging''}). Since $\bm{p}$ remains fixed across all the frames, if $N$ is sufficiently large, the simple average results in a proxy of the illumination pattern, blurred by the PSF: 
\begin{equation}
    \langle\{\bm{D}\}^N\rangle\sim\langle\left\{\bm{\rho}_a\cdot \bm{p} \right\}\otimes\bm{h}\rangle = \bm{p} \otimes\bm{h} =\bm{p}^0
\label{eq:simple averaging}
\end{equation}
where the $\langle \cdot \rangle$ operation averages out the translating component of the non-blurred fluorescent signal, $\bm{f}=\bm{\rho}\cdot \bm{p}$.
Let us consider the process of ``image registration'', which realigns the acquired frames. The registered images are obtained as  
\begin{equation}
    \overline{\mathbf{D}}_a = T(\mathbf{D}_a,-\bm{\Delta}_a).
\label{eq:registration}
\end{equation}
where $-\bm{\Delta}_a=[-\Delta x_a, -\Delta y_a]$ represents the inverse of the known translations applied to the object.
After registration, all fluorescence features are aligned, while the illumination pattern shifts according to the inverse of the original translations. 
Therefore, given the known displacement amplitudes, it is possible to compute an average of the registered frames. This results in a blurred proxy of the object:
\begin{equation}
    \langle{\{\bm{D}}\}^N\rangle^r\sim\langle\left\{\bm{\rho}\cdot \bm{p}_{-a}\right\}\otimes\bm{h}\rangle = \bm{\rho} \otimes\bm{h}=\bm{\rho}^0.
\label{eq:registered averaging}
\end{equation}
where the $\langle \cdot \rangle^r$ operator denotes the combined process of registration and averaging (\textit{``registered averaging''}). Fig.~\ref{fig:averaging} shows an example of simple averaging (top panel) and registered averaging (bottom panel) computed for a stack of 200 images.
Given the acquired data and the known displacements, these two operations effectively allow us to demultiplex $\bm{\rho}$ from $\bm{p}$, a task that is generally not straightforward since they always appear multiplied in the data (see Eq.~(\ref{eq:data equation}) and Eq.~(\ref{eq:data equation C-Sim})).
Here, we propose to integrate both simple and registered averaging in a RL-like iterative algorithm, thus enabling to extend one of the most common image reconstruction methods to the blind illumination architecture. 

\subsection{The generalized Richardson-Lucy algorithm}
\label{sec:algorithm}
The RL algorithm \cite{richardson1972bayesian,lucy1974iterative} improves image quality through its iterative refinement process, effective handling of Poisson noise, maintenance of non-negative pixel values, reliance on accurate PSF modeling, and provides competitive performance compared to other deconvolution methods (see e.g. \cite{makarkin2021}, for a recent review). 
While initially designed for single-image processing and flat illumination, the method has been progressively adapted to accommodate various experimental conditions, being also applied to SIM (see e.g. \cite{ingaramo2014,strohl2015,chakrova2016}).
The standard RL is based on the iterative equation:
\begin{equation}
    \bm{f}^{i+1} = \bm{f}^i \cdot \left\{\frac{\bm{D}}{\bm{f}^{i} \otimes \bm{h}} \otimes \bm{h^{T}}  \right\} 
\label{eq:R-L}
\end{equation}
where the index $i$ indicates the $i$-th iteration and $\bm{h^{T}}$ is the transpose of the PSF. 
We propose a generalization of the RL algorithm that exploits the simple averaging (Eq.~(\ref{eq:simple averaging})) and the registered averaging (Eq.~(\ref{eq:registered averaging})) of a stack of $a=1, ..., N$ images to retrieve, separately, $\bm{\rho}$ and $\bm{p}$. It is summarized by the following equations:
\begin{align}
    &\bm{\rho}^i_a = T(\bm{\rho}^i,\bm{\Delta}_a) 
    \label{eq:translation operator 2}\\
    &\bm{f}^{i+1}_a = (\bm{\rho}^i_a\cdot\bm{p}^i ) \cdot \left\{\frac{\bm{D}_a}{ (\bm{\rho}^i_a\cdot\bm{p}^i )\otimes \bm{h}} \otimes \bm{h^{T}} \right\} 
    \label{eq:gen R-L}\\
    &\langle\{\bm{f}^{i+1}\}^N\rangle = \bm{p}^{i+1}
    \label{eq:p retrieval}\\
    &\langle\{\bm{f}^{i+1}\}^N\rangle^r = \bm{\rho}^{i+1}
    \label{eq:rho retrieval}
\end{align}
This generalized-RL algorithm (Gen-RL) performs with the following workflow:
\begin{enumerate}

    \item The initial guesses $\bm{p}^0$ and $\bm{\rho}^0$ are derived, respectively, from Eq.~(\ref{eq:simple averaging}) and Eq.~(\ref{eq:registered averaging})

    \item An initial estimate of the translated object for each acquisition, $\bm{\rho}^0_a$, is obtained by applying the translation operator to $\bm{\rho}^0$ (Eq.~(\ref{eq:translation operator 2}))

    \item The non-blurred fluorescent signal guess, $\bm{f}^{i+1}_a$, is retrieved for each acquisition $a$ by performing a single RL iteration (Eq.~(\ref{eq:gen R-L})) 

    \item The updated guesses $\bm{p}^{i+1}$ and $\bm{\rho}^{i+1}$ are obtained by the simple averaging (Eq.~(\ref{eq:p retrieval})) and the registered averaging (Eq.~(\ref{eq:rho retrieval})), respectively, of all the updated $\bm{f}^{i+1}_a$.

\end{enumerate}
For reader's convenience, we illustrate this workflow in the diagram shown in Fig.~\ref{fig:flow_diagram}. An example of pseudocode for the algorithm is reported in Appendix~A, while a Matlab implementation is shown in Code 1 \cite{Gen-RL}.
\begin{figure}[H]
    \centering
    \includegraphics[width=9cm]{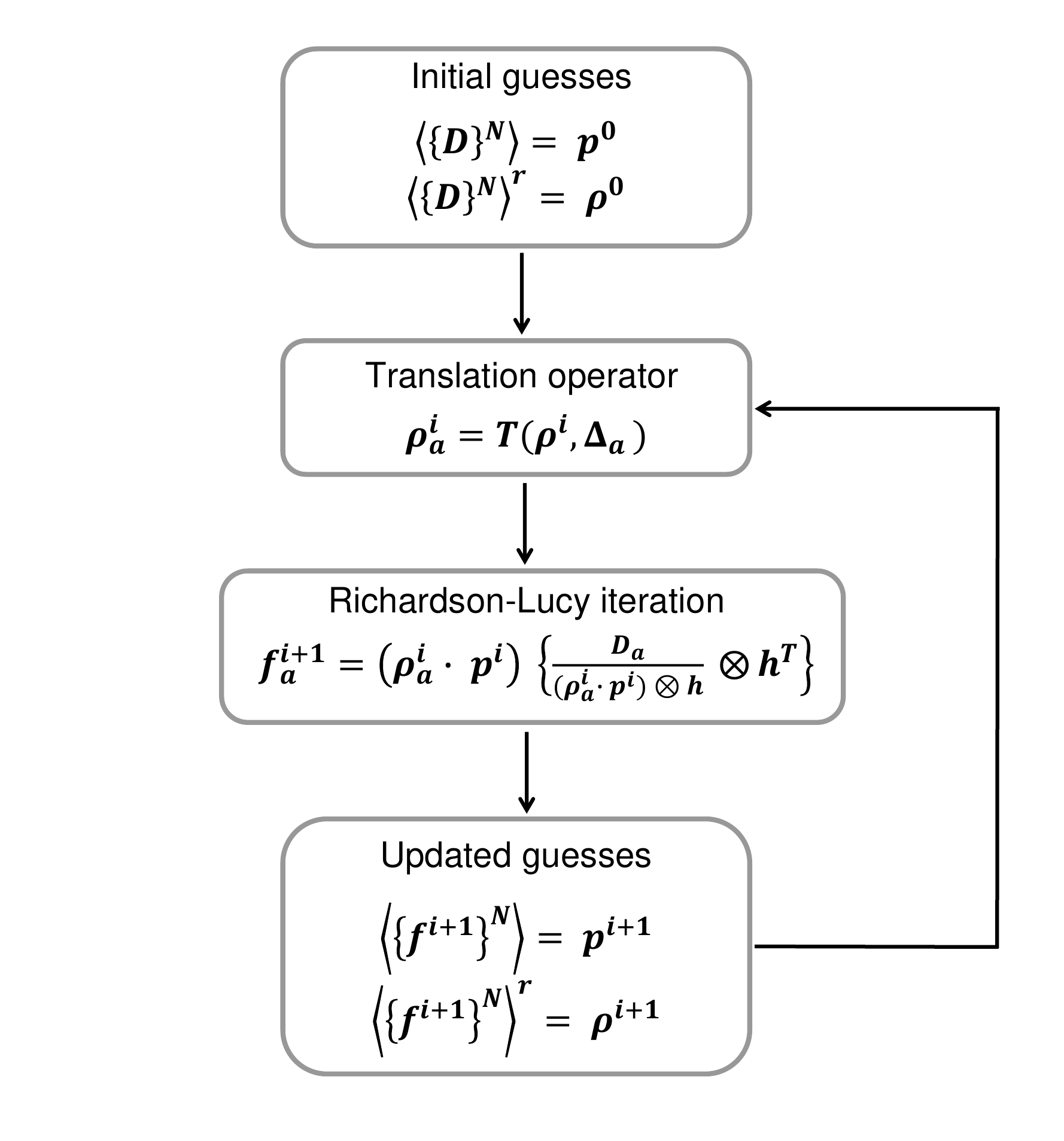}
    \caption{Example of a flow diagram for the Gen-RL algorithm. The notation follows that introduced in the text.}
    \label{fig:flow_diagram}
\end{figure}

\subsection{Experimental setup}
\label{sec:exp setup}
\begin{figure*}[htbp]
    \centering
    \includegraphics[width=\textwidth]{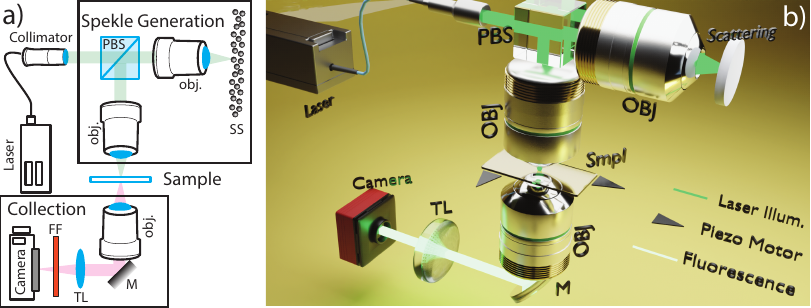}
    \caption{Experimental setup. Panel a) shows a schematic representation of the optical setup employed for the data acquisition. Panel b) shows a computer graphic representation of the same setup.}
    \label{fig:Setup}
\end{figure*}

We tested the Gen-RL algorithm and compared it to the one applied in C-SIM using the experimental setup illustrated in Fig.~\ref{fig:Setup}. In the speckle generation module, coherent light from a laser source (Oxious LCX-532-200 laser, $532\,nm$ wavelenght, $200\,mW $ maximum power), is delivered to the speckle generator module via a multimode fiber. The light is first collimated and then focused onto a strongly scattering medium. This medium is created by coating a standard aluminium mirror with a $\sim$ 100\,$\mu m$ layer of white paint. The back-scattered light is then reflected through a polarizing beam splitter onto a second objective, which generates a speckled illumination pattern on the sample to be investigated. After the sample, the collection module consists of an objective, a tube lens, a fluorescent filter, and a camera sensor. To ensure that the entire field of view is uniformly covered with speckle grains and that the speckle size matches the collection resolution across the entire field, we used three identical objectives (Olympus PLN 10X, Numerical Aperture=0.25). The fluorescent sample is translated using a closed-loop piezo actuator (NanoMax-TS, Thorlabs MDT630B/M). The voltage-to-pixel ratio is calibrated before the experiment. During acquisition, the applied voltage for the $x$ and $y$ axes is converted into pixel displacement and recorded. Details on the sample preparation are reported in Appendix~B.

\section{Results}

\subsection{Numerical tests} 
\label{sec:numerical tests}
\begin{figure}[H]
    \centering
    \includegraphics[width=10cm]{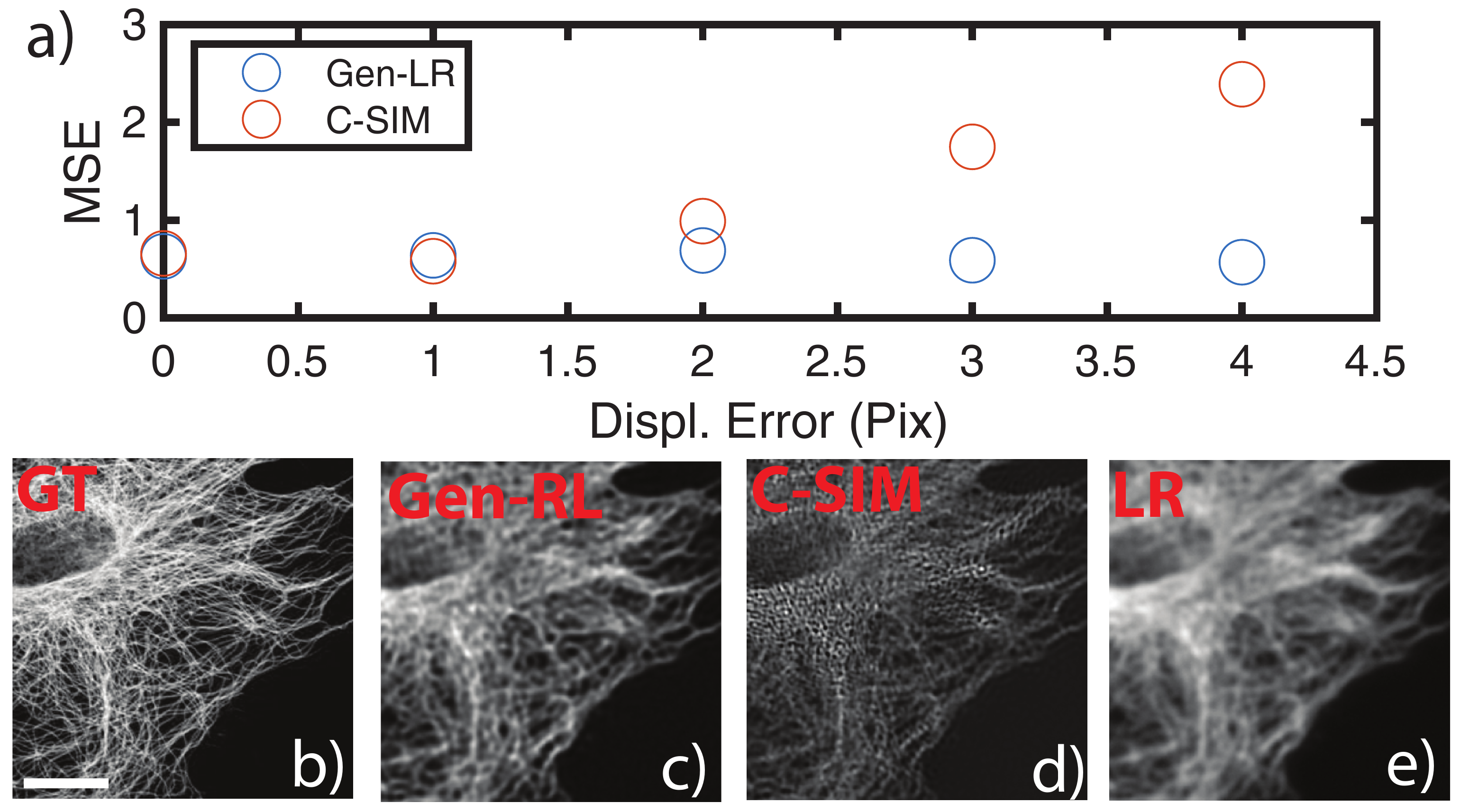}
    \caption{Comparison of the performances of Gen-RL and C-SIM algorithms with respect to displacement error and noise contribution. Panel a) shows the mean square error (MSE) between the image retrieved with Gen-RL (blue circles) and with C-SIM (orange circles) with respect to the ground-truth, along the displacement error in pixel units. Note that the MSE remains relatively stable due to the averaging behavior of the Gen-RL algorithm and the dominance of low-frequency components in the metric. Sensitivity to displacement errors becomes more evident in the resolution enhancement metric shown in Fig.~\ref{fig:res gain}. Panels b) – e) show, respectively, the ground-truth (GT), the image reconstructed with Gen-RL, C-SIM, and the low-resolution (LR) image obtained with registered averaging. Here, the GT is corrupted with a noise contribution of 1.5 counts and a displacement error of 3 pixels before applying the different algorithms. Image and illumination parameters are the same as in Fig.~\ref{fig:averaging}. Scale bar is $197\,\mu m$.}
    \label{fig:Displ Noise}
\end{figure}

\begin{figure}[ht]
    \centering
    \includegraphics[width=\textwidth]{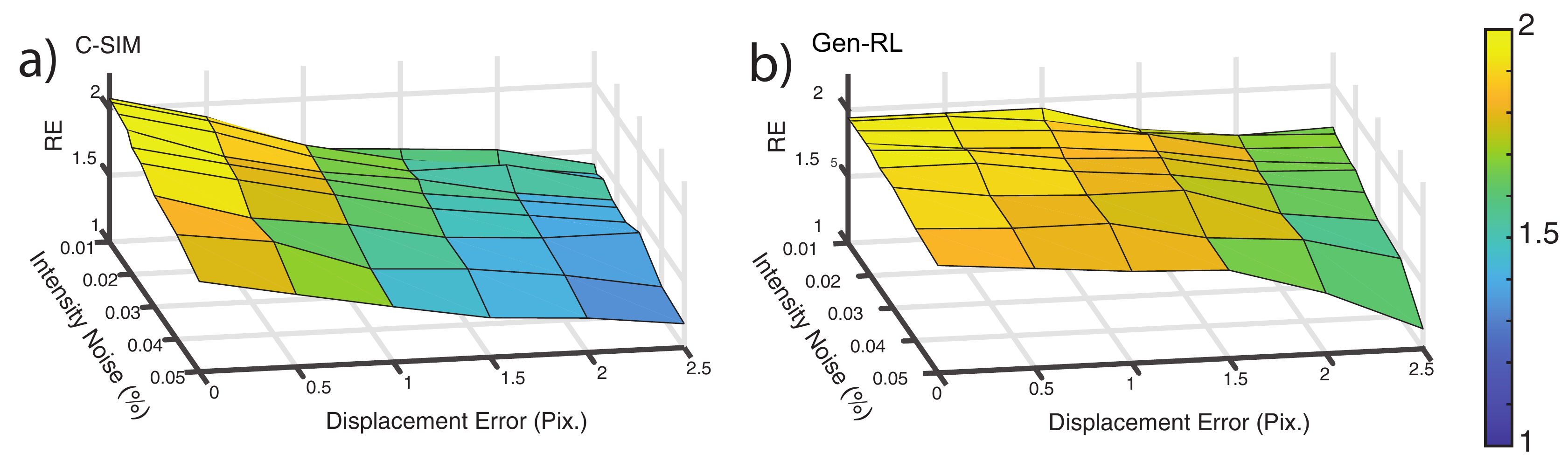}
    \caption{Resolution enhancement (RE), along noise intensity and displacement error, for C-SIM (panel a) and Gen-RL (panel b). The estimate is done by considering a stack of 256 images of a Siemens star illuminated with a simulated speckle pattern with a numerical aperture of 0.25.}
    \label{fig:res gain}
\end{figure}

\begin{figure}[ht]
    \centering
    \includegraphics[width=10cm]{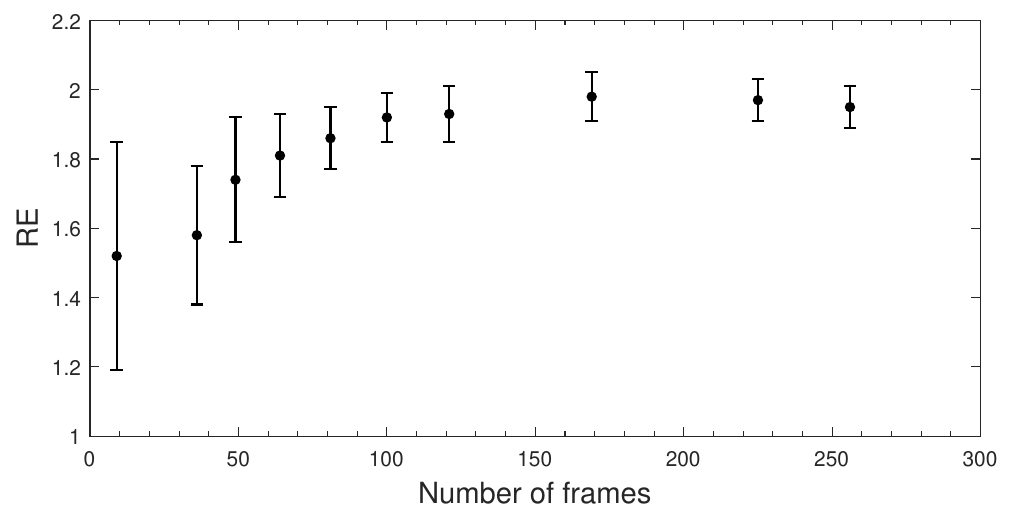}
    \caption{Resolution enhancement (RE) along number of acquired frames.}
    \label{fig:N_vs_RE}
\end{figure}

In the first C-SIM implementation, as presented in \cite{yeh2019computational,yeh2019speckle}, the guesses for the fluorescence distribution and the illumination pattern are updated resorting to gradients whose formulation is derived analytically without explicitly considering the noise contribution in the measurement (shot noise and readout noise) and the error on the positioning. 
On the other hand, the original RL algorithm was developed for the restoration of blurred images corrupted by noise, therefore it is capable to recover both a deblurred and denoised estimate of the fluorophores distribution. In this sense, the RL-based algorithm presented in this work is noise resilient (or noise ready).
In a laboratory configuration, additionally to the noise on the value of the intensity at each pixel, also the error on the actual sample displacement has to be considered. In fact, an inaccurate assessment of the translation between object and illumination results in a strongly aberrated or inefficient image retrieval.
In Fig.~\ref{fig:Displ Noise} we study these two contributions on the image reconstruction, showing that the Gen-RL algorithm surpasses C-SIM in terms of lower sensitivity to noise and displacement errors, as well as greater resilience to artifact generation.
Notably, the mean square error (MSE) for the Gen-RL algorithm remains relatively constant despite increasing displacement errors. This stability arises because the MSE is influenced by global image features and low-frequency components, which are less sensitive to fine displacement inaccuracies. Additionally, the Gen-RL algorithm implicitly compensates for small misalignments due to the averages performed over the shifted sample within the iterative procedure. As a result, while fine spatial resolution may be affected by these contributions (see Fig.~\ref{fig:res gain}), the global error as measured by MSE does not change significantly.
To quantitatively compare the two algorithms and better capture the effect of displacement on high-frequency information, we also estimate the resolution enhancement (RE) along the noise contamination and displacement error. We apply C-SIM and Gen-RL to a stack of 256 images, where the sample is represented by a standard Siemens star, and the illumination consists of a speckle pattern with a numerical aperture of 0.25. The RE is computed following the approach described in \cite{mudry2012structured}, and the results are shown in Fig.~\ref{fig:res gain}. For nearly noiseless data acquisitions, C-SIM achieves an RE of 2.076, while Gen-RL reaches 1.94. However, Gen-RL shows a key advantage: it maintains this enhancement nearly constant over a broader range of increasing noise strengths and displacement error, whereas RE rapidly decreases for C-SIM.
Additionally, in Fig.~\ref{fig:N_vs_RE}, we show RE as a function of the number of acquired frames. It is worth noting that, beyond the improvement in RE, increasing the number of frames also enhances image quality by reducing artifacts caused by inhomogeneous illumination of the sample.

\begin{figure}[H]
    \centering
    \includegraphics[width=9cm]{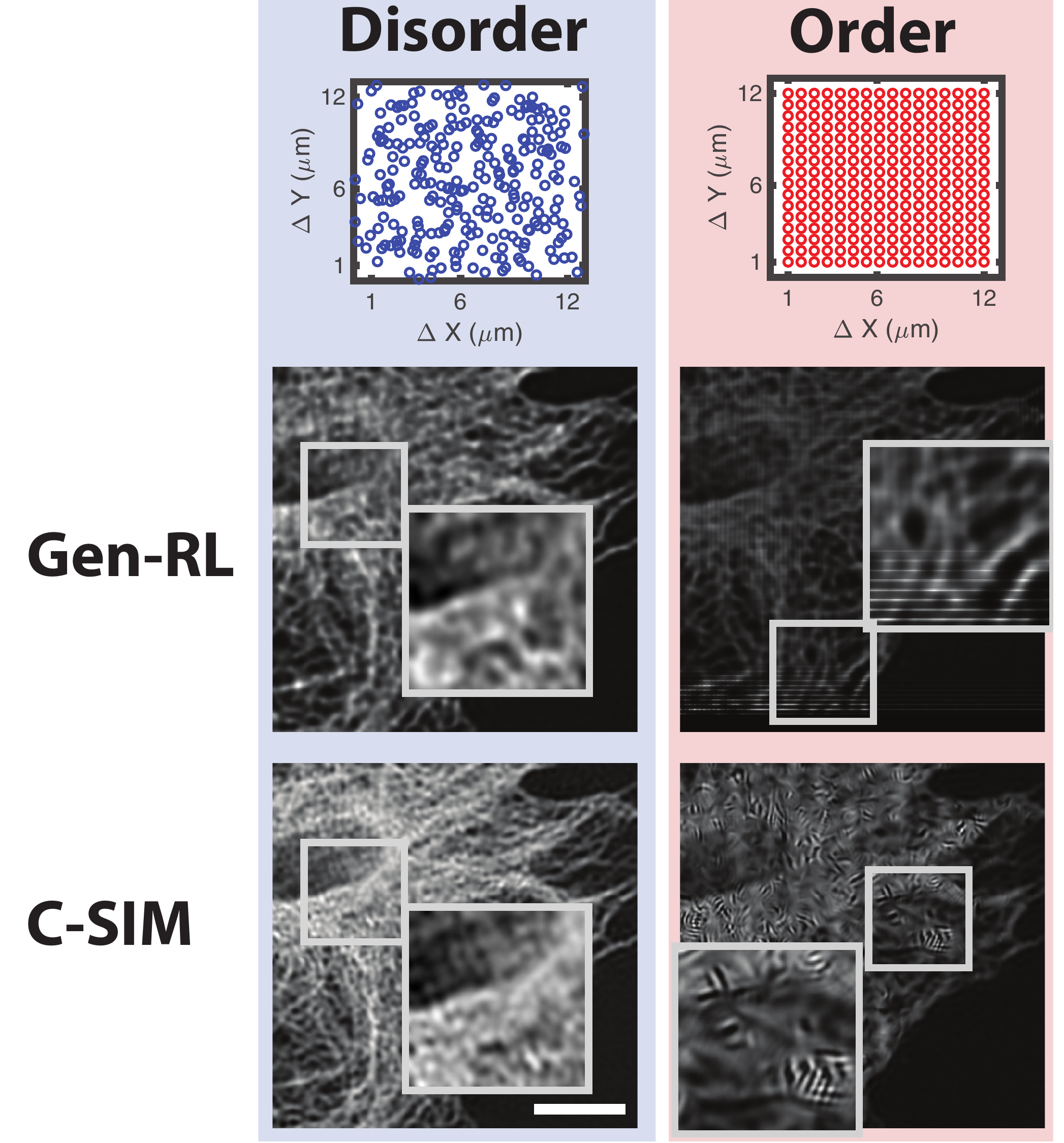}
    \caption{Impact of the displacement error on Gen-RL and C-SIM algorithms. \textit{Left}: image reconstructed using Gen-RL (middle panel) and C-SIM (bottom panel) after applying random translations (top panel) of the object. \textit{Right}: same cases as in the left, but considering ordered translations of the object. The insets in the middle and bottom panels highlight the artifacts that emerge in both algorithms due to the ordered arrangement of the translations. Scale bar is $197\,\mu m$.}
    \label{fig:Ord-Dis}
\end{figure}

We also analyse the impact of the specific realization of the displacement array--particularly the degree of ``order''--on the efficiency of both algorithms. Specifically, we compare a fully ordered displacement array, where displacement positions follow a structured pattern, with a completely disordered array, where displacements are randomly distributed. 
In our experimental setup, the ordered lattice of translations has a side length of $\mathcal{L}=13\,\mu m$, meaning the maximum displacement is approximately 10 times the instrumental PSF ($\sim 1.3\,\mu m$). The translation points are arranged in a square lattice, with a unit cell of size $\mathcal{S}=\mathcal{L}/N_p$, where $N_p$ is the number of points per dimension, resulting in a total of $N=N_p^2$ acquisitions/translations. In the randomized case, we introduce a $\pm 0.5\mathcal{S}$ displacement along both the $x$ and $y$ directions at each point. The displacement values are generated numerically using a uniform random number generator.
The results for both the Gen-RL and the C-SIM are shown in Fig.~\ref{fig:Ord-Dis}. It can be observed that the fully ordered  configuration produces artifacts in both approaches, whereas the fully disordered configuration is less prone to such artifacts. The onset of these artifacts is probably due to the fact that a set of ordered displacements may introduce ``beats'' at certain frequencies, which can result eventually in an overestimation of intensity in both $\bm{\rho}$ and $\bm{p}$. Conversely, a fully disordered displacement array helps to average out these beats, mitigating their impact. 

\subsection{Super-resolution characterization}
\label{sec:sr results}
\begin{figure*}[htbp]
    \centering
    \includegraphics[width=\textwidth]{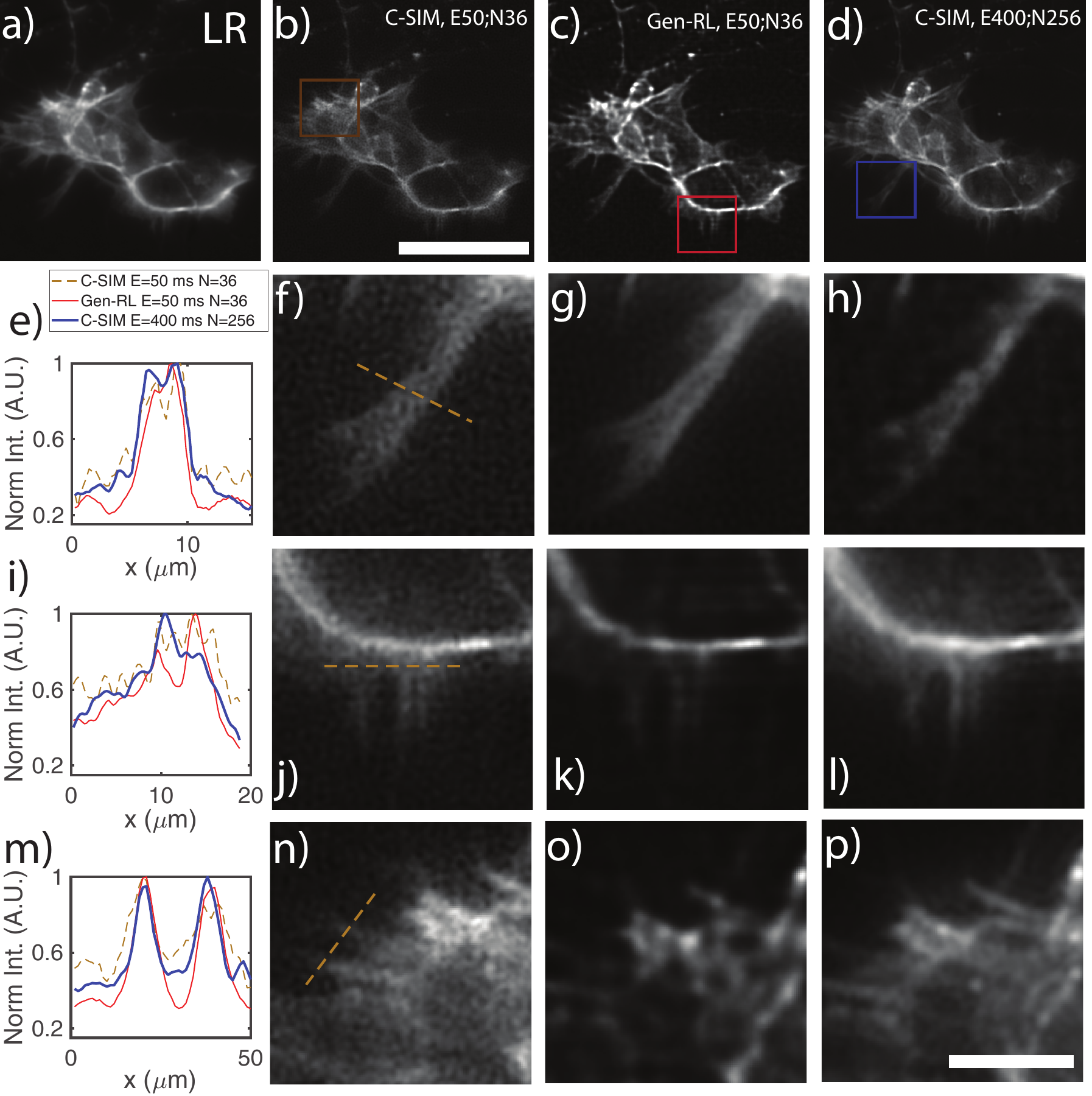}
    \caption{Results on a test sample. a) Low-resolution image of the sample obtained through registered averaging. b) - c) Reconstructed sample using C-SIM and Gen-RL, respectively, for M1 measurements. d) Reference results from M2 measurements reconstructed with C-SIM. f) - h) Zoomed-in results for the square region highlighted in blue in panel d), shown for the three cases above. j) - l) and n) - p) present the same comparison for the regions marked in red (panel c) and orange (panel b), respectively. e), i) and m) show the intensity profiles computed along the dashed lines for the three cases. Scale bars are $70\,\mu m$ in panels a) - d) and $17\,\mu m$ in panels f) - p).}
    \label{fig:Results}
\end{figure*}

The effectiveness of our experimental setup (see Section~\ref{sec:exp setup}) and the newly introduced SR algorithm is assessed based on measurements and image enhancement results, as illustrated in Fig.~\ref{fig:Results}. Here, the sample translations were performed following a disordered square lattice pattern, as described in Section~\ref{sec:numerical tests} (see the top-left panel of Fig.~\ref{fig:Ord-Dis}), which has demonstrated better performance in previous numerical tests. We performed two sets of measurements:
\begin{itemize}
    \item M1: low exposure time ($E=50\,ms$) and a small number of measurements ($N=36$)

    \item M2: high exposure time ($E=400\,ms$) and a large number of measurements ($N=256$)
\end{itemize}
Note that the total number of detected photons in M2 is 56 times greater than in M1, making M2 a nearly noiseless dataset. As proven in \cite{yeh2019computational}, C-SIM provides a resolution enhancement factor of 2 in low-noise regime, therefore we use M2 images reconstructed with C-SIM (shown in panel d in Fig.~\ref{fig:Results}) as a reference target. 
In Fig.~\ref{fig:Results}, panels b) and c) show the results by employing C-SIM and Gen-RL, respectively. Panels f) and g), j) and k), n) and o) report the same comparison for selected, zoomed-in subregions.
To further quantify the performance of the two algorithms, intensity profiles along the orange-dashed lines are reported in panels e), i) and m).
It is evident that the Gen-RL algorithm exhibits better performance in the short-exposure-time regime, highlighting its advantages in reducing residual noise contamination. However, a potential limitation of RL-like deconvolution is the risk of overfitting, which can occur when increasing the number of iterations in the presence of noise. Our analysis shows that $\sim 20$ iterations are sufficient to improve the error in the image reconstruction while preventing the introduction of additional artifacts.

\section{Discussion}
The Gen-RL method expands the class of translation-based SIM techniques applied under unknown illumination patterns. However, with respect to the previous C-SIM approach, it introduces several differences that affect reconstruction performance, robustness, and usability.
A primary distinction lies in the nature of the applied translations. C-SIM relies on ordered, grid-like translations of the illumination pattern, while Gen-RL uses disordered, random translations of the sample itself. This modification contributes to artifact suppression in the reconstruction, as demonstrated by our results.
On the algorithmic front, Gen-RL avoids gradient-descent optimization in favor of a generalization of the RL deconvolution. This adaptation results in an iterative update scheme that effectively handles the blind and shift-variant nature of the problem. By exploiting simple and registered averaging of the image stack, the method isolates the fluorescence signal from the unknown illumination, serving as a demultiplexing step prior to iterative refinement. This simplified framework not only improves usability but also enhances the method's resilience to experimental imperfections. Indeed, our comparative analysis shows that while C-SIM achieves slightly higher resolution enhancement under ideal, noise-free conditions, Gen-RL outperforms it across a broader range of noise levels and displacement errors. This robustness makes Gen-RL particularly suitable for real applications where acquisition conditions are less controlled. Finally, the method is implemented within a simplified and optimized optical design, enabling SR imaging across wide fields of view. This makes Gen-RL more practical and accessible for standard microscopy platforms.

\section{Conclusion}
We introduced a novel method for super-resolution imaging that employs SIM with an unknown illumination pattern, based on a generalization of the Richardson-Lucy (RL) algorithm for multi-frame acquisitions with translated objects. Our approach shows an increased noise resilience compared to previous techniques. The randomization of the translation coordinates, as opposed to a square (ordered) lattice, enables to further reduce artifacts and support the resolution enhancement over a large field of view. Our method exploits simple statistical properties of translated image stacks to demultiplex the fluorescence signal from the unknown illumination. We then apply an iterative RL-like algorithm to improve the performances in terms of image resolution and denoising. The experimental validation confirms that our technique is less sensitive to noise contributions while effectively minimizing artifact generation in the reconstruction process. We validated the method using an optimized experimental setup that simplifies the optical design. Our results demonstrate that this approach improves the compatibility of super-resolution imaging with standard wide-field microscopes.

\section*{Appendix A: Algorithm}
In the following, we provide an example of pseudocode for the proposed Gen-RL algorithm. A Matlab implementation can be found in Code 1 \cite{Gen-RL}.

\begin{algorithm}
\caption{The Gen-RL algorithm}

\begin{algorithmic}[1]
    \Require Raw images $\mathbf{D}_a$ (with $a=1,..., N$), translation amplitudes $\bm{\Delta}_a$, instrumental PSF $\bm{h}$
    \State initialize $\bm{p}^0 \gets \langle\{\bm{D}\}^N\rangle$
    \State initialize $\bm{\rho}^0 \gets \langle{\{\bm{D}}\}^N\rangle^r$
    \For {$i=0:i_{max}$} \Comment{$i_{max}$ is the number of iterations}
    \For {$a=1:N$}
    \State $\bm{\rho}^i_a \gets T(\bm{\rho}^i,\bm{\Delta}_a)$
    \State $\bm{f}^{i+1}_a \gets (\bm{\rho}^i_a\cdot\bm{p}^i ) \cdot \left\{\frac{\bm{D}_a}{ (\bm{\rho}^i_a\cdot\bm{p}^i )\otimes\,\bm{h}} \otimes \bm{h^{T}} \right\}$
    \EndFor
    \State $\bm{p}^{i+1} \gets \langle\{\bm{f}^{i+1}\}^N\rangle$
    \State $\bm{\rho}^{i+1} \gets \langle\{\bm{f}^{i+1}\}^N\rangle^r$
    \EndFor
\end{algorithmic}

\end{algorithm}

\section*{Appendix B: Cell preparation}
Asynchronous Human Retinal Pigment Epithelium cells (hTERT RPE-1) were fixed in 3.7\% formaldehyde in PBS for 10 minutes at room temperature (RT). They were then washed with 1M glycine for 15 minutes, permeabilized with 0.5\% Triton X-100 for 8 minutes, and blocked in 3\% BSA in PBS for 30 minutes. To stain filamentous actin (F-actin), the cells were incubated at RT for 1 hour with Phalloidin-Atto 532 (\#49429, 1:50, Sigma-Aldrich), a toxin that specifically binds to F-actin. Nuclei were counterstained with Hoechst reagent.

\begin{backmatter}
\bmsection{Funding}
MUR PRIN 2022 (2022CFP7RF  CUP: B53D23018560006). European Research Council, project ASTRA (grant agreement No. 855923). European Innovation Council, project ivBM-4PAP (grant agreement No. 101098989).

\bmsection{Acknowledgment}
This work was supported by MUR PRIN 2022 (2022CFP7RF  CUP: B53D23018560006 M.L.). This research was also funded by the D-Tails-IIT Joint Lab (to  M.L.) The research leading to these results was also supported by European Research Council through its Synergy grant program, project ASTRA (grant agreement No. 855923) and by European Innovation Council through its Pathfinder Open Program, project ivBM-4PAP (grant agreement No. 101098989).

\bmsection{Disclosures}
The authors declare no conflicts of interest.

\bmsection{Data Availability}
Data underlying the results presented in this paper are available on \href{https://opticapublishing.figshare.com/s/af5744ef69f8b1cf4d5f?file=53265773}{figshare}, Ref.\cite{Gen-RL}.

\end{backmatter}

%%%%%%%%%%%%%%%%%%%%%%% References %%%%%%%%%%%%%%%%%%%%%%%%%
%%%%%%%%%% If using BibTeX:
\bibliography{biblio}

\end{document}